\documentclass{amsart}
\usepackage{amsfonts,amsmath,graphicx}
\newcommand{\be}{\begin{equation}}
\newcommand{\ee}{\end{equation}}
\newcommand\re[1]{(\ref{#1})}

\begin{document}

\author{P. V\'an$^1$, C. Papenfuss$^2$, W. Muschik$^3$}
\title{Griffith cracks in the mesoscopic microcrack theory}
\markright{Mesoscopic microcracks}
\address{$^1$Technical University of Budapest\\
Department of Chemical Physics \\
1521 Budapest, Budafoki \'ut 8} \email{vpet@phyndi.fke.bme.hu}
\address{$^2$Technische Universit\"at Berlin\\
Institut f\"ur Mechanik, Institut f\"ur Theoretische Physik\\
Stra\ss e des 17. Juni\\10623 Berlin }
\address{$^3$Technische Universit\"at Berlin\\
Institut f\"ur Theoretische Physik\\
Hardenberg str. 36\\10623 Berlin }
\date{\today}
\begin{abstract}
The mesoscopic concept is applied to the description of
microcracked brittle materials. The mesoscopic equations are
solved in a special case when the microcracks are developing
according to the Rice-Griffith evolution law. The evolution of the
crack distribution function is investigated in case of simple
loading conditions and for two different initial crack
distribution functions. The time dependence of the average crack
length is calculated, too.
\end{abstract}


\maketitle

\section{Introduction}

There are several different approaches for the description of the
behavior of material with complex microstructure, e.g. granular
media, liquid crystals or damaged solid materials. One of the
basic approaches is the macroscopic one applying thermodynamic
concepts and principles to describe the microstructure (e.g.
internal thermodynamic variables). The other basic but different
approach is the statistical one where the microstructure is
considered as a statistical ensemble and the macroscopic variables
are calculated from the interactions between the microstructural
elements using the principles of statistical mechanics. Modelling
the microstructure tests the limits of applicability of both
theories and therefore the basic principles of our understanding
regarding the material behavior.

The microscopic elements of statistical cracking phenomena are the
single cracks. The form and validity of single crack evolution
equations is an important problem in statistical microcracking.
Beyond the well known classical results (see e.g.
\cite{Bro99b,Law89b}) there are new and promising approaches and
investigations in this field that should be considered. Langer and
his coworkers \cite{FalLan98a,LobLan98a,Lan00a} gave a microscopic
background of the internal variable approach of plasticity
\cite{Ric71a}. Using molecular dynamic simulations they found some
good microscopic candidates of the internal variables. Based on
these observations a so called {\em shear deformation zone} (STZ)
model was established, suggesting a dynamical equation of these
internal variables, including also special unilateral
requirements. Several inconsistencies of classical single crack
models can be explained using this theory. Another interesting
approach to single crack phenomena considers cracking as a kind of
nucleation problem, where elastic and conservation effects are
included in the nonlocal (coarse grained) phase field model
\cite{AraAta00a,KarAta01a,EasAta02a}, or the elastic effects are
included into a more direct nucleation approach
\cite{BreAta99a,BreAta00a,BreAta01a}. Both STZ and the phase field
approaches can be considered as phenomenological thermodynamic
theories.

On the other hand, partially independently of the previous
considerations microcracking is treated also from a statistical
point of view. There are two competing models in this respect.
Sornette and his coworkers considers cracking as a critical
phenomenon where rupture is a critical point. Their approach is
supported by experimental scaling observations, by numerical
simulations (see \cite{GluSor01a,AndAta97a} and the references
therein). Other observations and simulations suggest that the
rupture of the material is better considered as a first order
phase transition and the cracking as spinodal nucleation
\cite{ZapAta99a}. Scaling calculations based on the previously
mentioned nucleation models partially support that approach. From
the point of view of nonequilibrium thermodynamics the background
of both the statistical and single crack models is the violation
of thermodynamic stability in the nonequilibrium state space. This
observation alone can unify several classical failure and fracture
criteria and explain experimental observations
\cite{Van01a1,VanVas01p}.

However, between pure phenomenological and statistical approaches
there is an independent family of theoretical descriptions that we
can call {\em hypercontinuum theories}. In these theories the
physical quantities are not simple macroscopic fields given on the
appropriate space-time (classical, special or general
relativistic), but one assumes that in addition to the space-time
variables they depend also on some mesoscopic variables
characterizing the microstructure of materials.  These theories
have almost as long history as the traditional macroscopic and
microscopic ones. At the beginning they were used to model
rotating internal particles in microcrystalline materials (see the
basic works of the Cosserat brothers and Mindlin
\cite{CosCos09b,Min64a}), nowadays generalized continua are used
to model e.g. granular media \cite{ForSab98a,For98a,ForAta00a},
liquid crystals \cite{ChrAta99a,EhrHes95a}, ferrofluids
\cite{IlgKro02a,IlgAta02a}, polymer solutions
\cite{KroSel95a,KroHes93a}, but one can meet hypercontinuum in
general relativity, too \cite{ObuTre93a,Obu96a} (see also
\cite{Eri99b} and the references therein) .

Hypercontinuum theories represent models of the material where one
can get information from below the macroscopic but above the
microscopic level. That gives the basic motivation and modelling
idea to the so called mesoscopic theory, that uses both
macroscopic, continuum and microscopic, statistical concepts
constructing the basic equations of motion for the hypercontinuum.
This approach was successfully applied for the description of
liquid crystals (see \cite{MusAta00a2} and the references
therein).

Recently a mesoscopic theory was suggested for a special class of
damaged materials, for microcracked brittle solid media
\cite{VanAta00a}. In this paper we will see, that considering the
peculiar properties of microcracking simplifies the mesoscopic
equations well enough to make them solvable with relatively little
effort. This simplification results in a generalized Liouville
equation that we will solve considering a simple evolution
equation for the microcracks in case of two different initial
crack distribution functions. We applied Rice-Griffith microscopic
dynamic equations as the simplest unilateral microdynamics that is
exactly (analytically) solvable on the statistical level. The
explicit expression of the microcrack distribution function makes
possible to investigate the relevance of some damage parameter and
its respective evolution equation and the role of microscopic
unilateral constraints.

\section{Basic concepts of mesoscopic microcracking}

In this section we give a short summary of the mesoscopic
microcrack theory and the related equations governing the
development of crack distribution function. For a more detailed
treatment we refer to \cite{VanAta00a}.

 In a mesoscopic theory of a microcracked continuum our
quantities are interpreted on the {\em directional space} that
includes the ${\bf l}$ length of the microcracks and the ${\bf x},
t$ space-time coordinates. The cracks are considered as penny
shaped surfaces with area $\pi l^2$ and orientation ${\bf n}$.
These two quantities characterizing the crack are put together as
a vector  ${\bf l} =l {\bf n}$, where $l$ is the crack radius. The
microstructure consists of this surface vector of the cracks. This
vector is an axial one, that we do not consider as an
antisymmetric tensor but for the sake of convenience we use normal
polar vectors with a symmetry requirement on their number density
function:
$$
N({\bf l},{\bf x}, t) = N(-{\bf l},{\bf x}, t).
$$

The basic equations of the mesoscopic theory are the mesoscopic
balances of the fundamental physical quantities. Due to the
enlarged configurational space, the mesoscopic velocities and the
mesoscopic material current densities have six components and can
be decomposed into a 'normal' part (with respect to the position
variable) and a 'mesoscopic' or 'directional' part. The
macroscopic number density of microcracks of any size and
orientation at position ${\bf x}$ and time $t$. $\bar{N}$ can be
introduced with an integration over the directional part of the
mesoscopic space \cite{PapAta00p}: $ \bar{N}({\bf x}, t) :=
\langle N({\bf l},{\bf x}, t)\rangle := \frac{1}{2}
\int_{\mathbb{R}^3 } N({\bf l},{\bf x}, t) {\rm d}V_l.$ Now the
balance of the mesoscopic number density, the generalized
continuity equation can be written as:
\begin{equation}
\frac{\partial N}{\partial t} + \nabla\cdot (N {\bf v}) +
\nabla_{l}\cdot (N {\bf v}_l) = \sigma_n.
\label{Nmass_bal}\end{equation}

Here ${\bf v}$ is the translational velocity, the space part of
the mesoscopic velocity and ${\bf v}_l$ is the 'directional' part.
$\nabla_l$ denotes a derivation with respect to the variable ${\bf
l}$ and as a part of the mesoscopic divergence represents physical
requirement that the number of crack with a given length at a
given material element can be changed by changing the length of
some smaller or larger cracks. $\cdot$ dot stands for the
divergence, the source term $\sigma_n$ characterizes the creation
and coalescence of microcracks.

We introduce a distribution function $f$ as the probability
density of a crack having a particular length and orientation as
follows
\begin{equation}
f(l, {\bf n}, {\bf x} ,t) =\frac{ N(l, {\bf n}, {\bf x}, t)}{
\bar{N}({\bf x},t)}
\end{equation}

From the mesoscopic balance of number density follows the equation
determining the time development of the crack distribution
function is
\begin{equation}
\frac{\partial f}{\partial t} + \nabla\cdot (f {\bf v}) +
    \nabla_l\cdot (f {\bf v}_l) =
\frac{-f}{\bar{N}} \left(\frac{\partial}{\partial t} +
    {{\bf v}}\cdot\nabla\right)   \bar{N}+ \sigma_n
\label{crdist_bal}\end{equation}

In addition we have balance equations for the mesoscopic densities
of all other extensive quantities e.g. one can give the mesoscopic
balances of mass, momentum, angular momentum and energy
\cite{MusAta01a}. The macroscopic  balances can be introduced with
averaging the corresponding physical quantities over the
additional, enlarged part of their domain, using the previously
introduced directional probability distribution function.

We are interested here in the evolution of crack lengths and
damage is described in terms of the crack length distribution
function. However, crack orientation was introduced as an
additional mesoscopic variable together with crack length, and
there may be an arbitrary distribution of crack orientations.
Special cases of orientation distributions are: an isotropic
distribution of all orientations, or, as the other extreme case
parallel orientations of all cracks. Very often we will meet cases
in between: the cracks are not oriented exactly parallel, but
there is a preferred orientation. In these cases the orientation
distribution can be rotation symmetric (uniaxial) or not
(biaxial). This classification in terms of the orientation
distribution is analogous to liquid crystals
\cite{MusAta95a1,MusAta00a3}, where the mesoscopic variable is the
orientation of elongated molecules. Here we want to consider only
averages over all crack orientations. We introduce the crack
length distribution function as the average over all crack
orientations:
\begin{equation}
f(l,{\bf x},t) = \frac{1}{4\pi}\int_{S^2}f({\bf l},{\bf x},t)d^2n
\quad .
\end{equation}
Averaging of the differential equation (\ref{crdist_bal}) over
orientations results in the equation of motion for the crack
length distribution (\ref{eq_mored}). (For more details see
\cite{PapAta03a}.) It is the same form of differential equation
for any orientation distribution. Only the value of the crack
length change velocity $v_l$ depends on the orientation
distribution via the effective stress acting on cracks of
different orientations. Therefore we have only a parametric
dependence of the differential equation to be solved for the crack
length distribution on the orientation distribution, and the same
form of solutions is expected for any orientation distribution.

The distribution functions are normalized:
\begin{equation}
\int_{S^2} \int_0^{\infty} f( {\bf l}, {\bf x},t) l^2 dl d^2n =1
\qquad \forall {\bf x},t
\end{equation} and
\begin{equation}
\int_0^{\infty} f(  l, {\bf x},t) l^2 dl  =1 \qquad \forall {\bf
x},t \quad .
\end{equation}

\section{Solution of the mesoscopic equations}

There are essentially two methods to gain information from the
basic equations of the mesoscopic theory. The first one, that was
used mainly for liquid crystals, is to calculate a moment series
expansion of the equations. Using this expansion we can introduce
a hierarchy of macroscopic internal variables together with their
evolution equations. This can be useful also in our special case,
the nature of macroscopic damage parameters and the properties of
their evolution equations is an old and strongly discussed problem
in continuum damage mechanics. However, now we will prefer another
method: we will solve the mesoscopic equations in simple cases to
get direct information how the distribution function of
microcracks evolves in time.

The system of mesoscopic balance equations given in the previous
section is too general for most of the practical problems and
materials in damage mechanics.  In case of brittle (rock or
ceramic) material the following simplifications are suitable
\begin{enumerate}
\item The base material does not have an internal spin, that is a
crack does not rotate independently from the base material. \item
There are no couple forces and couple stresses. \item There are no
external body forces (${\bf f} = 0$), \item The material is in
(local) mechanical equilibrium $(\dot{\bf v} =0)$. \item
\label{cond.vel} The translational velocity {\bf v} does not
depend on the crack size and orientation. It is the same for all
cracks, namely  it is equal to the velocity of the center of mass
of the surrounding continuum element (${\bf v}({\bf l},x,t) =
\bar{\bf v}(x,t)$). This does not imply restrictions on the length
change velocity. \item There is no creation of cracks. This is not
a crucial restriction. For example preexisting voids are
considered as microcracks of very small size. In the progress of
damage they are growing, but their number is constant. (The voids
in the initial stage are ''counted'' as cracks.) \item There is no
crack coalescence. Together with the previous assumption we
postulate that there are no crack sources, i.e. the number of
cracks is constant in time. Let us remark, that the absence of
coalescence does not mean the absence of interaction. Interaction
can be considered in the microscopic dynamics as we will see
below.
\end{enumerate}

Let us remark that these assumptions are not necessary, more
general cases can be treated in the frame of our model. E.g. crack
creation and coalescence can be considered through a suitable
source term in \re{Nmass_bal}. However, in this paper we maintain
the theory as simple as possible to investigate some effects of
microscopic unilateral dynamics.

All the previous assumptions are reasonable in case of solid
materials. Considering these assumptions we find that the balance
of moment of momentum is trivial, the balance of momentum
simplifies to the condition of mechanical equilibrium:
$$
\nabla \cdot {\bf t} =0,
$$

\noindent where {\bf t} is the stress. Moreover at present we are
not interested in the changes of the mass density and the produced
energy during cracking, we will concentrate on the evolution of
the crack distribution function.

As a result of the conditions 5-7 the macroscopic number density
of microcracks does not change in a material element (only the
lengths can be changed) and there is no source term at the
mesoscopic level. Therefore there is no source term in the
differential equation for the crack distribution function
(\ref{crdist_bal}). Due to condition 4 choosing a comoving frame
results in a zero translational velocity and we get the following
simple equation:
\begin{equation}
\frac{\partial f}{\partial t} + \nabla_l\cdot (f  {\bf v}_l) = 0.
\label{eq_mo}
\end{equation}

Let us mention here that in this case the equation above has a
direct interpretation. From a pure statistical mechanical point of
view this is a continuity equation for the probability density of
microcracks having the "microscopic" dynamical equation
$$
\dot{\bf l} = {\bf v}_l({\bf l}).
$$

Here the point denotes the substantial (comoving) time derivative.
However, the special form of the differential equation for the
distribution function depends on the equation for the length
change velocity. This expression can involve a dependence on
orientations and lengths of surrounding cracks and also several
other forms of crack interaction. The equation (\ref{eq_mo}) is a
generalized form of the classical reversible Liouville equation of
statistical mechanics, determining the dynamics of the
distribution function $f$ with the help of a known deterministic
(!) microscopic dynamics \cite{Mac92b}. From this observation it
is clearly seen, that in the simplest non-interacting, reversible
case the mesoscopic theory is fully compatible with the basic
statistical theories and continuum theories, too.

\section{Rice-Griffith dynamics}

As an example we will use one of the simplest, but well known
dynamics that was given by Rice from general thermodynamic
considerations giving an irreversible thermodynamic background to
the energetic considerations of Griffith \cite{Gri24a,Ric78a}. The
condition is derived for a single crack under uniform tensile
loading parallel to the direction of the crack surface normal. The
main assumption in the construction of the dynamics is that the
Gibbs potential depends not only on the stress but also on the
crack length in the following way
\begin{equation}
G(\sigma,l) =
    \alpha l - \frac{\beta}{2} \sigma^2 l^2 - \frac{\gamma}{2}
    \sigma^2. \label{GibbsGR1}
\end{equation}

Here $\alpha$ is usually considered as a kind of surface energy,
$\beta$ is connected to the energy release rate in case of simple
geometric situations, $\gamma$ is the elastic coefficient, $l$ is
the length of the crack and $\sigma$ is the tensile stress
parallel to the crack normal. Let us remark here that similar
conditions can be given in case of compressive stresses with
modified material parameters (a compressive stress can result in a
tensile force on a crack surface for certain crack orientations).
The first term in (\ref{GibbsGR1}) is usually interpreted a
surface energy of the crack, the second as the extra elastic
energy of the surrounding continuum and the last one is the pure
elastic term in this simple case. The sign and Legendre
transformation conventions are according to Landau and Lifsic
\cite{LanLif81b}.

The parameters $\alpha$, $\beta$ and $\gamma$, the coefficients in
the Gibbs potential,  are effective parameters of the surrounding
continuum, including other cracks.

From this Gibbs potential we can derive the entropy production
$\sigma_s$ due to the crack extension in case of isothermal
conditions
$$
\sigma_s =\left(\frac{\partial^2G}{\partial \sigma^2
}\right)^{-1}\left(\frac{\partial G}{\partial \sigma} -
\epsilon\right)\dot{\epsilon} - \frac{\partial G}{\partial l}
\dot{l}.
$$

It is easy to identify thermodynamic forces and currents in the
expression. Moreover, we are interested only in the crack
propagation effects, therefore we will suppose that during the
propagation of the crack the material is in mechanical
equilibrium, because the crack propagation is considerably slower
than the deformation process. We conclude, that
\begin{equation}
 \epsilon =
    -\frac{\partial G}{\partial \sigma} =
    (\gamma +\beta l^2) \sigma,
\end{equation}

\noindent where $\epsilon$ is the deformation and we arrive to the
evolution law of crack propagation
$$
\dot{l} = -L \frac{\partial G}{\partial l} = -\alpha' + \beta' l
\sigma^2.
$$

Here $L$ is a positive kinetic coefficient (according to the
Second Law) and $\alpha' = L \alpha$, $\beta' = L \beta$. The
equilibrium solution of this equation gives the celebrated
Griffith condition \cite{Gri24a}. After simple stability
considerations one can see easily that the cracks start to grow if
the right hand side of the above differential equation is positive
(at least if we are investigating brittle rocks under room
temperature) expressing the unilateral nature of crack
propagation. If the healing of the cracks is excluded, then the
above evolution law can be written more exactly as
\begin{equation}
\dot{l} =
\begin{cases}
-\alpha' + \beta' l \sigma^2 &\text{if} \quad \alpha' \leq \beta'
l
    \sigma^2, \\
0 & \text{otherwise}.
\end{cases}
\label{RG1}\end{equation}

This evolution equation will be a core for the mesoscopic
dynamical investigation resulting in an explicit expression for
the mesoscopic velocity. The above unilateral dynamics will be
called Rice-Griffith dynamics.

Let us remark here that there is much more information that can be
gained from these thermodynamic considerations. For example as a
condition of thermodynamic stability we get an upper limit for the
crack length ($\frac{\gamma^2}{3 \beta} > l^2$) that one cannot
read from the single (\ref{RG1}) evolution equation without
considering the mechanical interaction. The role of thermodynamic
stability in the failure of real materials is investigated in more
detail in \cite{Van01a1,VanVas01p,Van01t}, where it is shown that
localization and critical damage can be considered as the loss
thermodynamic stability, and some empirical failure criteria (like
Griffith) can be understood well from a thermodynamic point of
view.

\section{Solutions of the Liouville-approximation}

Let us consider now a sample with tensile hydrostatic loading. The
mesoscopic variable in a given frame can be decomposed into the
crack length and the crack orientation ${\bf l}= (l, \Theta)$.
Similarly the mesoscopic velocity can be decomposed into a length
change part and an orientation change part, as ${\bf v}_l =
(v_l,\omega)$. According to assumption 1 and 4 the cracks cannot
move independently of the surrounding continuum, and let be the
material at rest ${\bf v}={\bf 0}$, therefore the orientational
change velocity is zero ($\omega = 0$). Here we will investigate
pure tensile loading conditions, when the crack surface normal are
parallel to the applied loading. Due to our simplifications a the
role of crack orientation changes can be neglected in the
development of the different order parameters \cite{PapAta03a}.

In the following we will investigate constant loading rate
conditions $\sigma = v_\sigma t$, $v_\sigma = const.$ common in
standard failure tests. Considering all of our assumptions
regarding the cracks  we can conclude that the crack density
distribution depends only on time and the length of the cracks.
The evolution equation (\ref{eq_mo}) in spherical coordinates and
in the mesoscopic space is the following
\begin{equation}
\frac{\partial f}{\partial t}(t,l) =
    -\frac{1}{l^2} \frac{\partial l^2 v_l(l,t) f(t,l)}{\partial l},
\label{eq_mored}
\end{equation}

\noindent where the mesoscopic speed, the crack length change
velocity $v_l$ is given by the help of the right hand side of the
equation (\ref{RG1}), that is $v_l = \dot{l}$.

Easy to check that the general solution of the first order partial
differential equation (\ref{eq_mored}) with the microscopic
dynamics (\ref{RG1}) is the following
$$ f(l,t) =
\begin{cases} l^{-2} e^{-\frac{\beta'
    v_\sigma^2}{3}t^3} F\left(l e^{-\frac{\beta' v_\sigma^2}{3}t^3} +
    \frac{\alpha'}{(9 \beta' v_\sigma^2)^{1/3}}
        \Gamma(1/3,0, \frac{\beta' v_\sigma^2}{3}t^3) \right)
& \text{if}\quad \alpha' \leq \beta'v_\sigma^2 l t^2, \\
f(l,0) & \text{otherwise}.
\end{cases}
$$

Here $\Gamma$ is the generalized incomplete Gamma function defined
by $\Gamma(a,b,c) = \Gamma_i(a,c)-\Gamma_i(b,c)$, where $\Gamma_i$
denotes the incomplete Gamma function. $F$ is an arbitrary
function to be determined from the initial conditions. One can
observe that there is no stationary solution of (\ref{eq_mored}),
but the crack distribution with a finite support does not change
under a definite loading. A solution with arbitrary initial
conditions cannot be given analytically because of the non-healing
microcracks, the non differentiable right hand side of
(\ref{RG1}). As a consequence of this non-differentiability, the
solutions of the partial differential equation can be splitted
into two parts connected by a moving boundary. Smaller and smaller
cracks start to grow arriving at the moving boundary.  The
arbitrary function $F$ should be determined in such a way that the
final distribution function will be continuous.

In the following we introduce dimensionless variables with
measuring the time in units of $\hat{t} = v_\sigma^{-2/3}
\beta^{-1/3}$, the length in $\hat{l} = \alpha v_\sigma^{-2/3}
\beta^{-1/3}$ and the mass $\hat{m} = \alpha v_\sigma^{-5/3}
\beta^{-4/3}$ as the crack length $l$, the stress $\sigma$, the
stress change velocity $v_\sigma$ and the material parameters
$\alpha'$ and $\beta'$ are measured in SI units. The dimensionless
microdynamic equation (\ref{RG1}) does not contain material
parameters (e.g. can be written as $v_\sigma=1$, $\alpha'=1$ and
$\beta'=1$ in (\ref{RG1})). In the following all the variables are
understood as dimensionless.

The solution of this equation is calculated numerically in case of
two different initial length distribution functions. First an
exponential initial crack length distribution function was
considered
$$
f(l,0) = l^{-2} \frac{e^{-\frac{l}{\delta}}}{\delta}.
$$

\begin{center}\begin{figure}
\hfill\includegraphics[height=7.7cm]{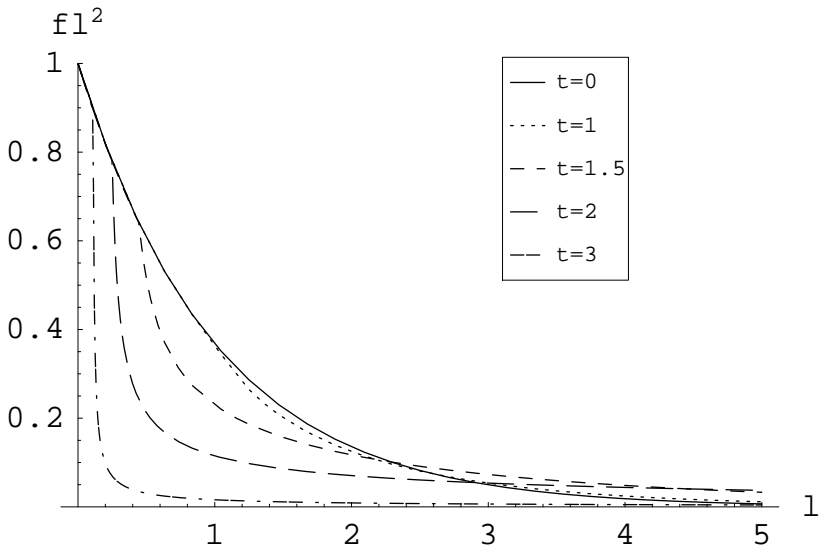} \caption{The
crack length distribution function at different instants.
Exponential initial condition. }
\label{fig2}\end{figure}\end{center}

You can see the solution with the parameter value $\delta=1$ at
different instants on figure \ref{fig2}. On the vertical axis the
function $l^2 f(l,t)$ is drawn (because $f$ multiplied with the
spherical weight has the direct probability interpretation). One
can observe, that the breaking point of the initial exponential
distribution moves to the left, according to Rice-Griffith
dynamics.

\begin{center}\begin{figure}
\includegraphics[height=10cm]{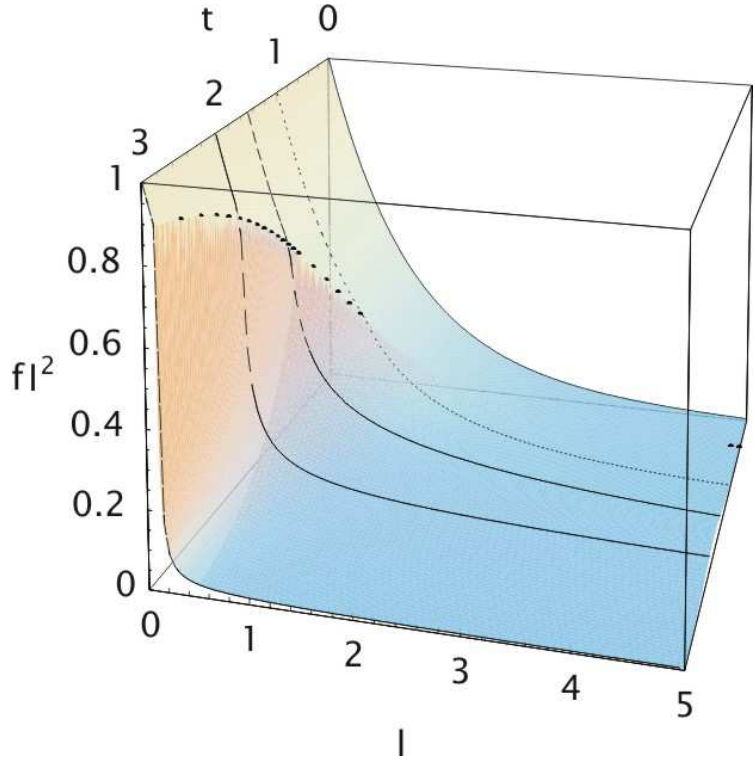}
\caption{The time development of the crack length probability
distribution. Exponential initial condition. }
\label{fig3}\end{figure}\end{center}

In figure \ref{fig3} we can observe the full time and crack length
development of the distribution function. The projection of the
thick pointed line, where the surface is broken, onto the $l-t$
plane gives the Griffith condition, considering the time
dependence of stress. The other lines show the distribution
function at the instants on figure 2.

As a more realistic example, that considers an upper limit of the
possible crack lengths we treat a bounded uniform initial crack
distribution function, too:
\begin{equation*}
f(l,0) =
\begin{cases}
\frac{l^{-2}}{l_c} & \text{if}\quad l<l_c\\
0 & \text{otherwise}.
\end{cases}
\end{equation*}

\begin{center}\begin{figure}
\includegraphics[height=7.7cm]{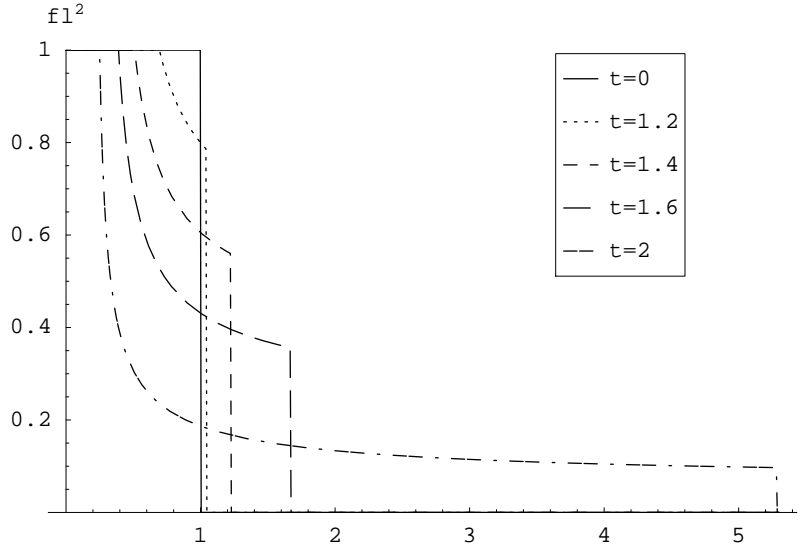}
\caption{The crack length distribution function at different
instants. Stepwise initial condition.}
\label{fig4}\end{figure}\end{center}

\begin{figure}
\includegraphics[height=6cm]{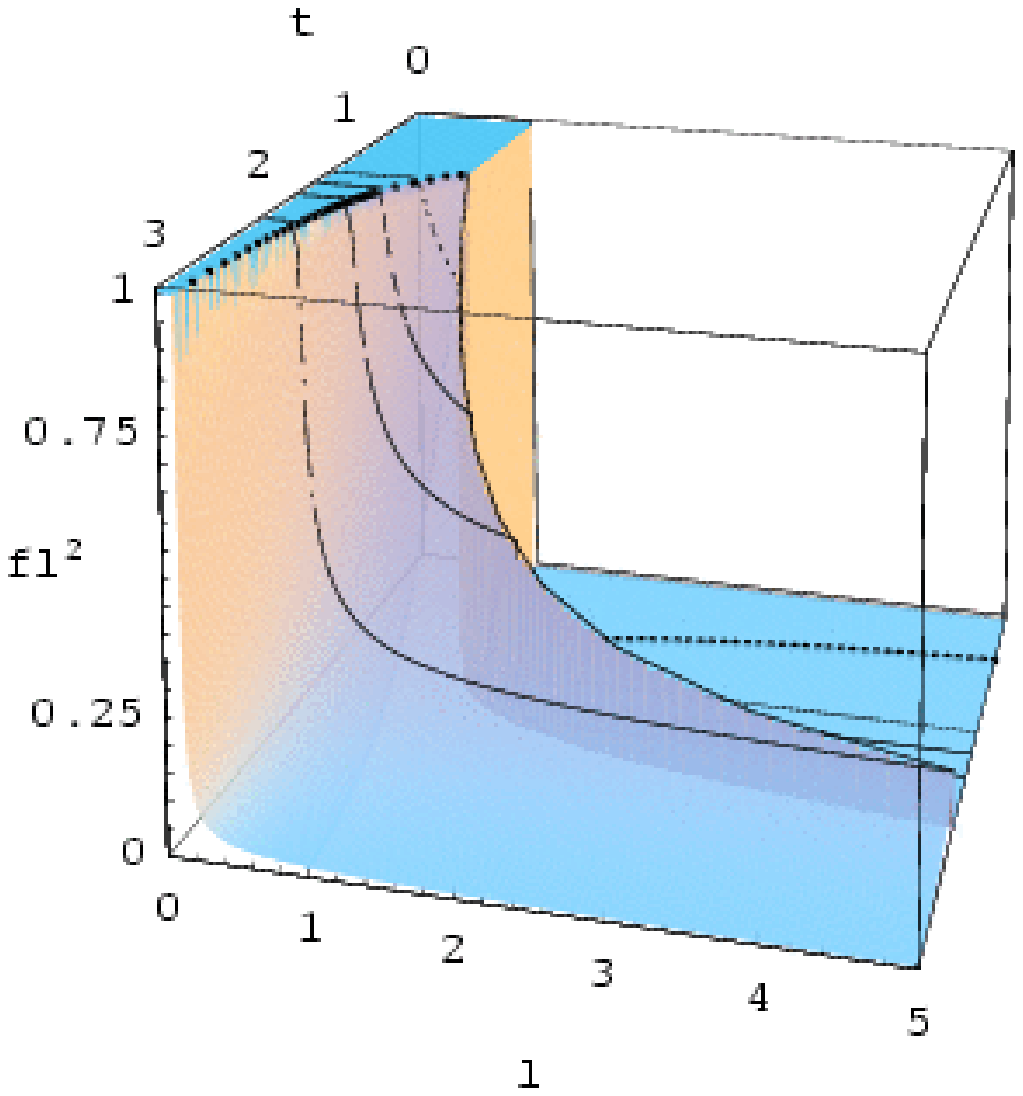}\hfill
\includegraphics[height=6cm]{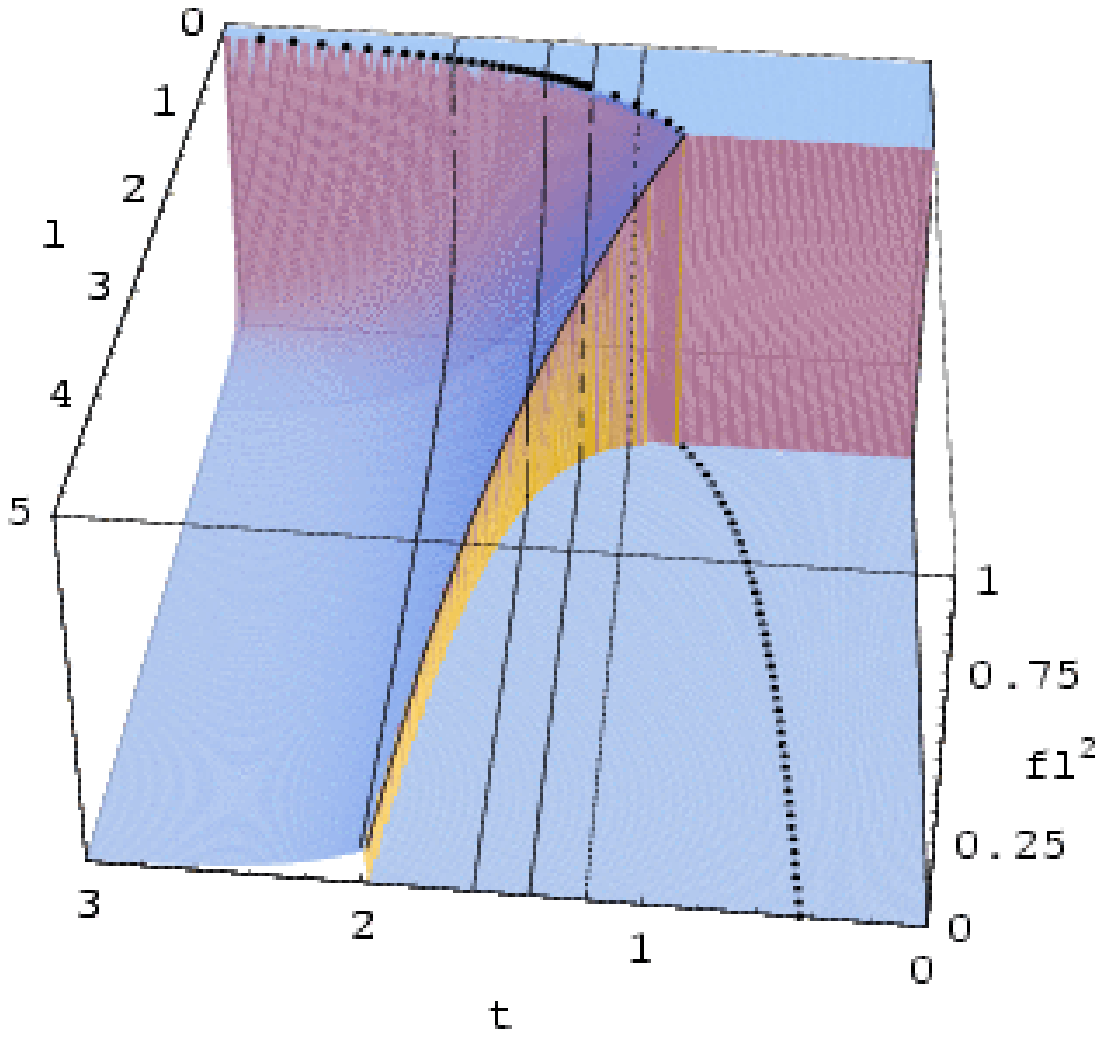}
\caption{Different views of the time development of the crack
length probability distribution. Stepwise initial condition. }
\label{fig6}\end{figure}

In the following we choose $l_c =1$. In figure \ref{fig4} one can
see the crack length distribution at different instants. Figure
\ref{fig6} shows together the time and length dependence. The
pointed line and its projection onto the $l-t$ plane is the
Griffith condition, that is more apparent on the backward view.
The thin lines are the crack distribution functions at the same
instants as in figure 4.

Here the resulted solution is discontinuous, due to the
discontinuous initial condition. The initial discontinuity, the
largest cracks, move to larger crack sizes with increasing time.

In \cite{PapAta03a} we suggest several possible macroscopic
internal variables from a mesoscopic point of view. Different
order and damage parameters are introduced and the possible
advantages and disadvantages are discussed. Here we will
investigate one of the most remarkable candidates, the average
crack length and calculate its time dependence in case of the
treated two initial crack length distributions. The average crack
length will be called in the following as {\em damage} and defined
as
$$
D(t) = <l> = \int_0^\infty l f(l,t)l^2 {\rm d}l = \int_0^\infty
l^3 f(l,t){\rm d}l.
$$

The notion of an order parameter is chosen  in analogy to the
mesoscopic theory of liquid crystals, where the analogously
defined orientational moments of the distribution function, the
alignment tensors, are the order parameters in a Ginzburg-Landau
theory. There the gradient of the order parameter contributes to
the free energy density in the form of the Frank elastic energy.
Here, in the mesoscopic theory of damage, we did not take into
account any spatial gradients at all, and therefore there is no
direct comparison to a Ginzburg-Landau theory. Definitely the
damage parameter introduced here is different from the order
parameter, the displacement field in phase field models. The
spatial structure of the crack is out of the scope of the
mesoscopic theory, and we are dealing only with microcracks
smaller than the typical dimension of the continuum element.  On
the other hand the onset of damage has been interpreted as a loss
of thermodynamic stability with the help of a free energy
depending on a damage parameter density \cite{Van01a1} . The
inclusion of spatial inhomogeneities of the damage parameter, and
the formulation of an extension of a Ginzburg-Landau type theory
of damage (with a mesoscopic background) is an outstanding problem
for the future.

\begin{center}
\begin{figure}
\hfill\includegraphics[height=7cm]{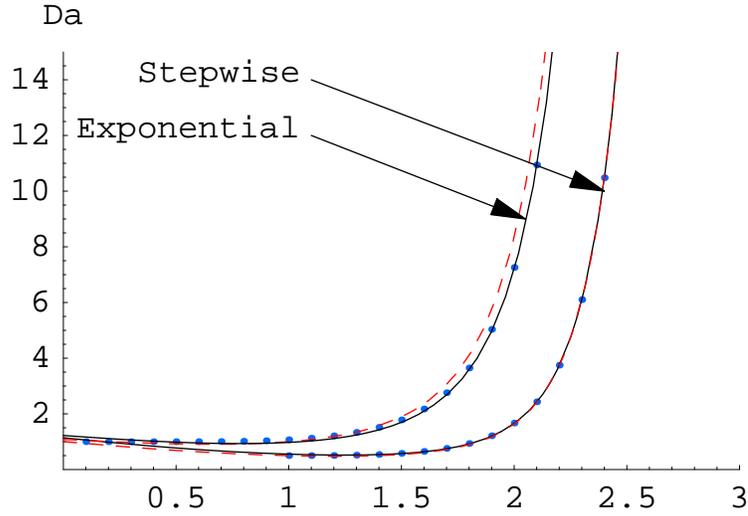}\hfill
\caption{Time dependence of damage, in case of exponential and
stepwise initial distribution. The dots are related to the
averaged distribution function, the dashed and continuous lines
are the lines calculated from the macrodynamic equations.}
\label{fig7}\end{figure}
\end{center}

\begin{center}
\begin{figure}
\hfill\includegraphics[height=7cm]{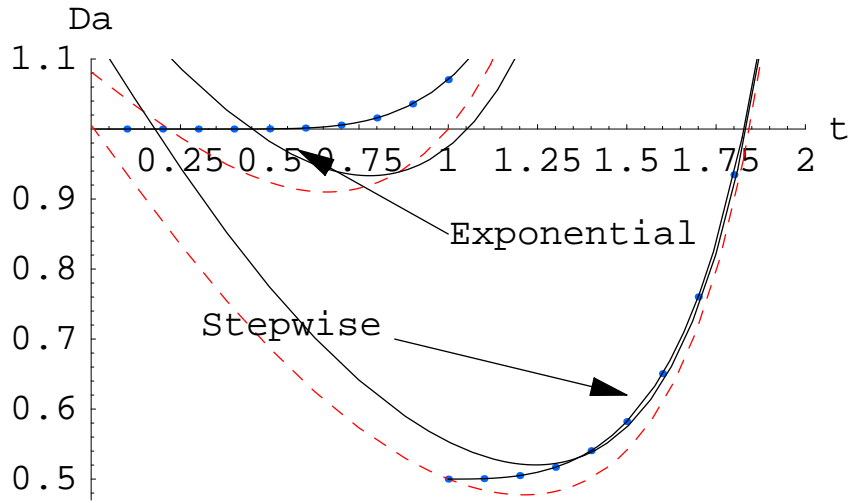}\hfill
\caption{Time dependence of damage, in case of exponential and
stepwise initial distribution. (Enlarged initial part of figure
\ref{fig7})} \label{fig8}\end{figure}
\end{center}

After calculating the time dependence of the damage (the dots on
figure \ref{fig7} and \ref{fig8}) we can observe that time
development is similar to the solution of the microscopic dynamic
equation. Therefore we can investigate, whether the damage
development is determined by a Rice-Griffith dynamical equation.
More properly we suppose that the evolution equation that
determines the damage dynamics is the following
\begin{equation}
\dot{D}(t) = -a + b t^2 D(t). \label{avdyn}\end{equation}

One can get these equations by averaging the mesoscopic dynamic
equations without considering the unilateral property of the
microdynamics. The above equation is the only reasonable candidate
to describe the macroscopic dynamics of the averaged crack length.

To compare the exact and the suggested macrodynamics we fitted the
solutions of the above equation with the exact solution to
determine the parameters of the damage dynamics and the dependence
on the initial conditions. In both cases we considered the
possibility, that the damage starts from the initial value
determined from the initial crack distribution (dashed lines on
figure \ref{fig7} and \ref{fig8}) or the initial damage belongs to
the parameters to be determined (continuous lines on figure
\ref{fig7} and \ref{fig8}). The results are summarized in the
table.

\begin{center}\begin{tabular}{|l|c|c|c|c|}\hline
       & $t_0$ & $D(t_0)$ & $ a $ & $ b $ \\ \hline\hline
Exponential  &  0  & 1       & 0.395 & 1 \\\hline
         &  0  & 1.2226 & 0.5681  & 1 \\\hline
Stepwise     &  1  & 0.5       & 0.7006  & 1 \\\hline
         &  1  & 0.5534 & 0.8061  & 1 \\\hline
\end{tabular}\label{tab1}\end{center}
\vskip 0.1in

The parameters were determined by an asymptotic approximation
method considering more and more sample points. We can observe
that the $\beta'=1$ parameter of the microscopic crack dynamics is
the same as the analogous $b$ in case of damage dynamics, however
the $\alpha'=1$ and $a$ parameters (the "surface energy" times the
dynamic coefficient) are smaller for both initial distributions.
The averaged macroscopic equations especially in case of fitted
initial conditions give remarkable well approximation for larger
times as one can see e.g. on \ref{fig7}. However, at the beginning
when the number of "frozen" cracks is large, the damage
developments are qualitatively different (see figure \ref{fig8}).
One can observe a kind of "pseudo healing", the damage curves
calculated from the macroscopic considerations start to decrease,
instead of the expected increasing. This effect is due to the
unilateral microdynamics. The averaged distribution function is
monotonously increasing but the suggested approximate macrodynamic
equation at the beginning start to decrease. The decreasing
depends on the initial distribution. The initial part of the
curves related to the stepwise distribution is an extrapolation in
both figures to show that an unilateral macrodynamic equation
would not improve the situation.

\section{Discussion and final remarks}

In this paper the equations of mesoscopic microcracking were
solved in case of simple cracking processes. We have seen that
under reasonable assumptions the seemingly complex mesoscopic
balances are reduced to a solvable first order partial
differential equation giving a kind of Liouville-approximation of
the mesoscopic theory. There we could get the mesoscopic
velocities introducing some constitutive assumptions based on
known single microcrack dynamics. The partial differential
equation has an independent statistical physical interpretation
and can be useful to describe the evolution of microstructure
without the underlying mesoscopic theory, too. Moreover, it gives
a sure basis of the internal variable calculations, giving
particular interpretations of the different internal variables and
clues regarding their dynamics, the changes of the elastic moduli
or the stress distribution without any further physical
assumptions.

The development of the average crack length was investigated in
detail and we have got a macroscopic dynamics that was almost the
same as for single cracks with the remarkable property that the
corresponding parameters of the damage evolution equation strongly
depend on the initial crack distribution. Moreover, the
statistical calculations resulted in a new qualitative phenomena:
the average crack length was decreasing at the beginning,
seemingly contradicting to the unilateral microdynamics of the
microcracks. That calculations give some insight into the
applicability of internal variable approaches in nonequilibrium
situations. The unilateral microscopic conditions alone (without
quenched disorder) resulted in a macroscopic dynamics of the
damage (internal variables) that proved to be a good approximation
of  to the microscopic dynamics, but with parameters strongly
depending on the initial crack distribution.

Our assumptions regarding the microscopic, single crack
propagation evolution equations can be extended to far more
difficult cases, for example we can consider more general loading
conditions, a microscopic dynamics with terminal crack velocity,
too.

\section{Acknowledgements}

This research was supported by OTKA T034715 and T034603, by the
DAAD, the DFG and by the VISHAY Company, 95100 Selb, Germany. Most
of the figures and some calculations were prepared and done by
Mathematica 4.2. Thanks for the referees for their valuable and
deep observations.

\end{document}